\documentclass[%
 reprint,
 amsmath,amssymb,
 aps,
]{revtex4-1}

\usepackage[usenames, dvipsnames]{color}
\usepackage{graphicx}
\usepackage{dcolumn}
\usepackage{bm}

\begin{document}

\preprint{APS/123-QED}

\title{Asymptotic Behavior and Limiting Distribution of Quantum Walk on Cycles with General U(2) Coin by Reduced Characteristic Matrix Method \\}

\author{Majid Moradi}
  \email{majid.moradi@shahroodut.ac.ir}
  \email{majidofficial@gmail.com}
\author{Mostafa Annabestani}%
 \email{annabestani@shahroodut.ac.ir}
\affiliation{ Faculty of Physics, Shahrood University of Technology
}%

\date{\today}

\begin{abstract}
We calculated reduced density characteristic matrix (RDCM) for quantum walk on cycles (QWC) to study asymptotic properties of the most general form of quantum walk on cycles with general $U(2)$ coin operator. As  an example, entanglement temperature for general initial state has been calculated and compared to previous results. Also, we have modified RDCM to derive analytical expression for general form of limiting distribution (LD).

\end{abstract}

\pacs{Valid PACS appear here}
\maketitle


\section{\label{sec:level1}Introduction}
Quantum Walk (QW) has been introduced by Aharonov et al. \cite{YAharonov1993} and has been widely used in different algorithms for solving problems \cite{Ambainis2007, Buhrman2006, Magniez2005, Farhi2007, Reichardt2009} . This wide usage encouraged more people to study this field.

There are two general variants of QWs known as discrete-time QW \cite{Watrous2001} and continuous-time QW \cite{Farhi1998}. In continuous-time QW the walk can be defined directly on position space \cite{Farhi1998}, while in discrete-time QW a coin operator defines the direction of movement for the particle \cite{Ambainis2005}. It has been shown that due to coin degree of freedom, discrete-time QW is more powerful than continuous-time QW \cite{Ambainis2005}. In one-dimensional QW \cite{Ambainis2001} there are two directions for the particle to move along, so the coin operator is a unitary $2 \times 2$ operator. One of the most used coins is the Hadamard coin, first used by \cite{Ambainis2001}.

People studied different aspects of QWs. Some considered different topologies for position space, for example QW on line \cite{Ambainis2001}, QW on planes \cite{Mackay2002}, quantum walk on M\"obius strip \cite{Moradi2017} and QW on hypercubes \cite{Moore2002}. Also QWs on cycles (QWC) has been defined \cite{Aharonov2001, Bednarska2003}. Some others considered asymptotic aspects such as hitting time \cite{Kempe2005} or mixing time \cite{Ambainis2001} and limiting distribution on cycles \cite{Aharonov2001, Bednarska2003}. Some others focused on entangled and non-local initial states \cite{Abal2006}. A few studied the effects of environment in QW and investigated decoherence \cite{Annabestani2010}.

Romanelli et al. focused on asymptotic aspect of QW and defined thermodynamic quantities such as entanglement temperature from asymptotic reduced density matrix \cite{Romanelli2014}. The concept also has been extended to quantum walk on cycles \cite{Diaz2016}.
In this paper, we have used method of \cite{Annabestani2019} to find reduced density characteristic matrix (RDCM) for general form of quantum walk on cycles (QWC) and we also modified RDCM method to derive exact form of limiting distribution (LD) for general form of QWC. In order to verify our formalism, we have calculated some of asymptotic properties of QWC, including asymptotic entanglement temperature and LD for Hadamard walk, as well as LD for non-local initial states.

In section \ref{sec:level2} we first provide a quick review of quantum walk on cycles and after that we present two general formulas for asymptotic averaged density matrix and reduced density matrix. In section \ref{sec:level3}, we provide different examples for application of the general formulas in section \ref{sec:level2} in order to study: A) Asymptotic temperature, B) Limiting distribution with local initial state and C) Limiting distribution with non-local initial state.

\section{\label{sec:level2}Quantum Walk on Cycles}
In quantum walks (QW), the total Hilbert space is $\mathcal{H} = \mathcal{H}_{c} \otimes \mathcal{H}_{p}$, where $\mathcal{H}_{c}$ is the coin space and $\mathcal{H}_{P}$ is the position space. For QWs with two direction of movement (such as one-dimensional QW or quantum walk on cycles) the coin space is a 2D Hilbert space. The dimensions of position space can be infinite for one-dimensional QW, while equals number of nodes $N$ for quantum walk on cycles (QWC). So the state of the walker at each step can be shown as
\begin{equation}\label{IniitialState}
| \psi(t) \rangle = \sum\limits_{s = 0}^1 {\sum\limits_{j = 0}^{N - 1} {{a _{s,j}(t)} | {s,j} \rangle } }
\end{equation}

where $ | s \rangle$ and $ | j \rangle $ are the state of the walker in coin and position subspaces, respectively.
In QWC, the nodes are distributed on a circle, so the particle moves on a circular path and can reach the beginning point after accomplishing a complete round. Each step in QWC consists of a unitary $2 \times 2$ coin operator $\Gamma$ applying on particle's coin state and a subsequent shift operator $S$ applying on particles's position state. So, for the initial state introduced in Eq.\eqref{IniitialState} 
\begin{eqnarray}\label{UPsi}
| \psi (t+1) \rangle = U | \psi (t) \rangle =S ( \Gamma \otimes I_p) | \psi (t) \rangle
\end{eqnarray}
where $I_p$ is identity matrix on position space and $U$ is evolution operator. The most general form of $\Gamma$, as a $U(2)$ coin, is
\begin{eqnarray} \label{GeneralCoin}
\Gamma= e^{\frac{i \eta} {2}} \left[ {\begin{array}{*{20}{c}}
	{{e^{i\zeta }}\cos (\theta )}&{{e^{i\xi }}\sin (\theta )}\\
	{ - {e^{ - i\xi }}\sin (\theta )}&{{e^{ - i\zeta }}\cos (\theta )}
	\end{array}} \right]
\end{eqnarray}
without losing generality we can omit the phase $\eta$.

The shifting operator $S$ can be considered in the form of
\begin{equation}\label{ShiftOperator}
S = \sum\limits_{j = 0}^{N - 1} {\sum\limits_{s = 0}^1 {\left|s, {\left( {j + {{\left( { - 1} \right)}^s}} \right)mod \, N } \right\rangle \left\langle {s,j} \right|} }.
\end{equation}
$S$ has been defined for QWC in order to move the walker to the left/right according to $s=0,1$ respectively.

We follow the method introduced by \cite{Annabestani2019} to find characteristic matrix for QWC and show that this characteristic matrix can be useful to calculate limiting distribution of QWC.

By using discrete Fourier transformation
\begin{equation}\label{Fourier}
| {{\kappa _k}} \rangle  = \frac{1}{{\sqrt N }}\sum\limits_{n = 0}^{N - 1} {{e^{\frac{{2i\pi kn}}{N}}} | n \rangle } 
\end{equation}
we can transform the non-diagonal form of the evolution operator $U$ in position space into block-diagonal form in k-space. So for any $k$ there is a $\tilde{U}_k$ \cite{Portugal2013}:
\begin{eqnarray}\label{UK}
{\tilde U_k} = \left[ {\begin{array}{*{20}{c}}
	{{e^{ - i\omega }}}&0\\
	0&{{e^{i\omega }}}
	\end{array}} \right] \times \Gamma
\end{eqnarray}
where $\omega = \frac{2 \pi k}{N}$. So eigenvalues are the eigenvalues of each block (two eigenvalues for each block)
\begin{equation}\label{Eigenvalues}
\lambda_{k}^{(m)}= e^{\pm i \alpha} \,\,\, , \,\,\, \cos(\alpha)= \cos(\theta) \cos(\omega - \zeta)
\end{equation}

The eigenstates for $\tilde{U_k}$ are
\begin{equation}\label{Eigenstates}
\begin{array}{l}
\left| {{\lambda^{(0)} _k}} \right\rangle = \left[ {\begin{array}{*{20}{c}}
	{ - {e^{i\left( {\xi  - \omega } \right)}}\sin \left( \theta  \right)}\\
	{{e^{ - i\alpha }} - {e^{ - i\left( {\zeta  - \omega } \right)}}\cos \left( \theta  \right)}
	\end{array}} \right]\\ \\
\left| {{\lambda^{(1)} _{k}}} \right\rangle = \left[ {\begin{array}{*{20}{c}}
	{ - {e^{i\left( {\xi  - \omega } \right)}}\sin \left( \theta  \right)}\\
	{{e^{i\alpha }} - {e^{ - i\left( {\zeta  - \omega } \right)}}\cos \left( \theta  \right)}
	\end{array}} \,\, \, \, \right]
\end{array}
\end{equation}
By using the spectral decomposition of evolution operator in $k$ space
\begin{equation}\label{USpectralDecompositin}
\tilde{U}_k = \sum\limits_{k = 0}^{N - 1} {\sum\limits_{i = 0}^1 { \left| k \right\rangle \left\langle k \right| \otimes {\lambda ^{(i)}_k}\left| {{\lambda ^{(i)}_k}} \right\rangle \left\langle {{\lambda ^{(i)}_k}} \right| } } 
\end{equation}
On the other hand, for initial state $| \psi (0) \rangle$, after $t$ steps, the state of the particle is
\begin{equation}\label{Psit}
| \Psi(t) \rangle = U^{t} | \Psi (0) \rangle
\end{equation}
Using \eqref{UPsi}, \eqref{USpectralDecompositin} and \eqref{Psit}, one can write
\begin{equation}\label{PsiSpectral}
\left| {\Psi \left( t \right)} \right\rangle  = \left( {\sum\limits_{k = 0}^{N - 1} {\sum\limits_{i = 0}^1 {\lambda _k^{(i)t}\left| k \right\rangle \left\langle k \right| \otimes \left| {\lambda _k^{(i)}} \right\rangle \left\langle {\lambda _k^{(i)}} \right|} } } \right)\left| {\Psi \left( 0 \right)} \right\rangle
\end{equation}
\\
Since $\left| {\Psi \left( 0 \right)} \right\rangle$ has been defined on the total space coin+position we define amplitude of initial state in $k$ space as
\begin{equation} \label{KdotPsi}
\left\langle {k}
\mathrel{\left | {\vphantom {k {\Psi \left( 0 \right)}}}
	\right. \kern-\nulldelimiterspace}
{{\Psi \left( 0 \right)}} \right\rangle  = \left| {{\psi _k}} \right\rangle
\end{equation}
and rewrite \ref{PsiSpectral} as
\begin{equation}
\left| {\Psi \left( t \right)} \right\rangle  = \sum\limits_{k = 0}^{N - 1} {\sum\limits_{i = 0}^1 {\lambda _k^{(i)t}\left| k \right\rangle  \otimes \left| {\lambda _k^{(i)}} \right\rangle \left\langle {\lambda _k^{(i)}} \right|\left. {{\psi _k}} \right\rangle } }
\end{equation}
\\
Now we can use the state of the walker to form its density matrix at time $t$
\begin{eqnarray}\label{RhoCExtended}
\rho (t)  &=& \sum\limits_{k,k' = 0}^{N - 1} {\sum\limits_{i,j = 0}^1 {\lambda _k^{(i)t}\lambda _{k'}^{(j){t^*}} \left| {k} \right\rangle \left| {\lambda _k^{(i)}} \right\rangle \left\langle {\lambda _{k'}^{(j)}} \right|\left\langle {k'} \right|} } \nonumber \\
&\times& \left\langle {{\lambda _k^{(i)}}}
\mathrel{\left | {\vphantom {{\lambda _k^{(i)}} {{\psi _k}}}}
	\right. \kern-\nulldelimiterspace}
{{{\psi _k}}} \right\rangle \left\langle {{{\psi _{k'}}}}
\mathrel{\left | {\vphantom {{{\psi _k}} {\lambda _{k'}^{(j)}}}}
	\right. \kern-\nulldelimiterspace}
{{\lambda _{k'}^{(j)}}} \right\rangle 
\end{eqnarray}
\\
Since the eigenvalues of evolution operator are in exponential form, so always after a certain number of iterations the state of the walker will be repeated, thereupon its state does not converge \cite{Portugal2013}. That's why people study time averaged asymptotic properties.
\\
We can define time averaged density matrix as \cite{Aharonov2001} $\bar \rho = \frac{1}{T} \sum\limits_{t}^{T} \rho (t)$. Therefore from \eqref{RhoCExtended}
\begin{eqnarray}
\bar \rho &=& \frac{1}{T} \sum\limits_{t}^{T}\sum\limits_{k,k' = 0}^{N - 1} {\sum\limits_{i,j = 0}^1 {({\lambda _k^{(i)}\lambda _{k'}^{(j){*}}})^{t}\left| k \right\rangle \left| {\lambda _k^{(i)}} \right\rangle \left\langle {\lambda _{k'}^{(j)}} \right|\left\langle {k'} \right|} } \nonumber \\
&\times& \left\langle {{\lambda _k^{(i)}}}
\mathrel{\left | {\vphantom {{\lambda _k^{(i)}} {{\psi _k}}}}
	\right. \kern-\nulldelimiterspace}
{{{\psi _k}}} \right\rangle \left\langle {{{\psi _k{'}}}}
\mathrel{\left | {\vphantom {{{\psi _k}} {\lambda _{k'}^{(j)}}}}
	\right. \kern-\nulldelimiterspace}
{{\lambda _{k'}^{(j)}}} \right\rangle 
\end{eqnarray}
when $T \rightarrow \infty$, the only time-dependent term i.e. $ \frac{1}{T}\sum\limits_t {({\lambda _k^{(i)}\lambda _{k'}^{(j){*}}})^{t}}$ can be simplified as below \cite{Aharonov2001}
\begin{equation}\label{LargeTLambdaLambda}
\mathop {\lim }\limits_{T \to \infty } \frac{1}{T}\sum\limits_t {(\lambda _k^{(i)}{\lambda _k{^{(j)}}^*)^t} = \left\{ {\begin{array}{*{20}{c}}
		1&{}&{{\lambda _{k} ^{(i)}} = {\lambda _{k '} ^{(j)}}}\\
		{}&{}&{}\\
		0&{}&{{\lambda _{k} ^{(i)}} \neq {\lambda _{k '} ^{(j)}}}
		\end{array}} \right.} 
\end{equation}
In fact we can define $\tilde{\rho}$ as an asymptotic form of average density matrix as follows
\begin{align} \label{LmtngDensMat}
\tilde{\rho}  &= \sum\limits_{k,k' = 0}^{N - 1} {\sum\limits_{\substack{i,j = 0 \\ \lambda _k^{(i)} = \lambda _{k'}^{(j)}}}^1 {\left| k \right\rangle \left| {\lambda _k^{(i)}} \right\rangle \left\langle {\lambda _{k'}^{(j)}} \right|\left\langle {k'} \right|} } \nonumber \\
& \times \left\langle {{\lambda _k^{(i)}}}
\mathrel{\left | {\vphantom {{\lambda _k^{(i)}} {{\psi _k}}}}
	\right. \kern-\nulldelimiterspace}
{{{\psi _k}}} \right\rangle \left\langle {{{\psi _k{'}}}}
\mathrel{\left | {\vphantom {{{\psi _k}} {\lambda _{k'}^{(j)}}}}
	\right. \kern-\nulldelimiterspace}
{{\lambda _{k'}^{(j)}}} \right\rangle 
\end{align}
Note that inner summation has the constraint $\lambda _k^{(i)} = \lambda _{k'}^{(j)}$.

\begin{figure}
	\includegraphics[scale=.3]{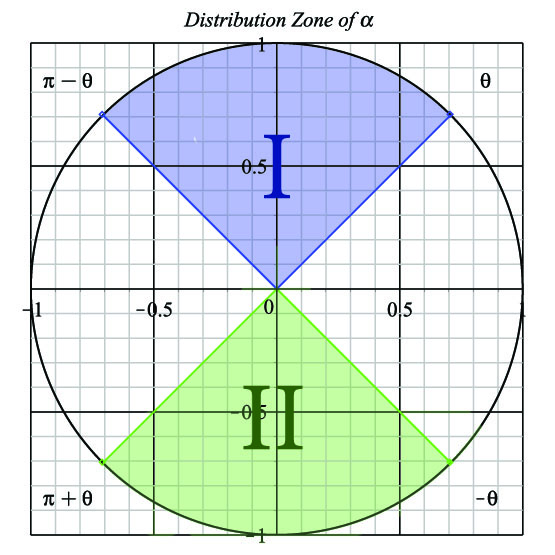}
	\caption{Distribution zones of $\alpha$}\label{EigenZones}
\end{figure}

\begin{figure}
	\includegraphics[scale=.3]{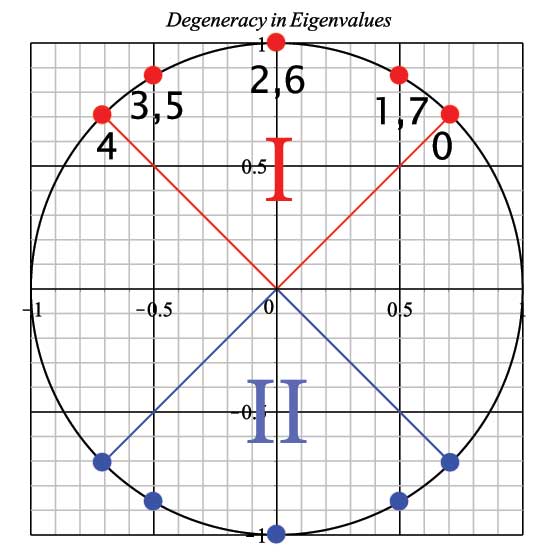}
	\includegraphics[scale=.3]{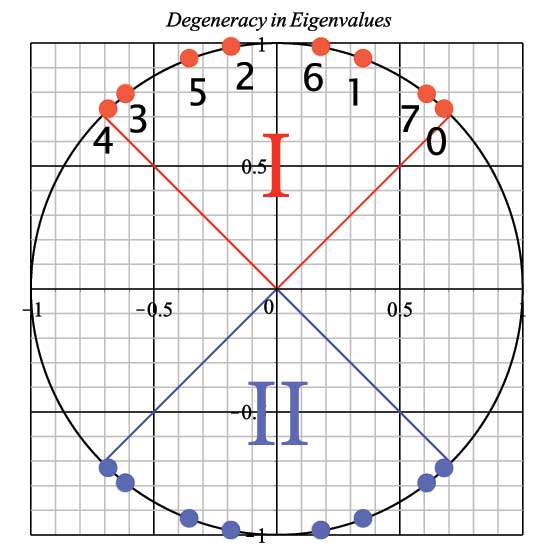}
	\caption{Degeneracy in eigenvalues for (up) $N=8$ and $\zeta =0$ and (down) $N=8$ an $\zeta =\frac{\pi}{3}$.}\label{DegEig}
\end{figure}

Since $-cos(\theta) \leq cos(\alpha) \leq cos(\theta)$, the eigenvalues are in two different zones (see Fig. \ref{EigenZones}), i.e. $e^{i\alpha}$ belongs to zone I and $e^{-i\alpha}$ belongs to zone II and therefore the eigenvalues from different zones ($i \neq j$) do not contribute in \eqref{LmtngDensMat}, therefore
\begin{eqnarray} \label{LmtngDensityMat2}
\tilde{\rho}  &=& \sum\limits_{k,k' = 0}^{N - 1} {\sideset{}{'}\sum\limits_{i = 0}^{1} {\left| k \right\rangle \left| {\lambda _k^{(i)}} \right\rangle \left\langle {\lambda _{k'}^{(i)}} \right|\left\langle {k'} \right|} } \nonumber \\
&\times& \left\langle {{\lambda _k^{(i)}}}
\mathrel{\left | {\vphantom {{\lambda _k^{(i)}} {{\psi _k}}}}
	\right. \kern-\nulldelimiterspace}
{{{\psi _k}}} \right\rangle \left\langle {{{\psi _k{'}}}}
\mathrel{\left | {\vphantom {{{\psi _k}} {\lambda _{k'}^{(i)}}}}
	\right. \kern-\nulldelimiterspace}
{{\lambda _{k'}^{(i)}}} \right\rangle 
\end{eqnarray}
where $\Sigma '$ means the summation is just over the terms in which ${\lambda _k^{(i)}} = {\lambda _{k'}^{(i)}}$. Note that degeneracy of ${\lambda _k^{(i)}} = {\lambda _{k'}^{(i)}}$ depends on $\zeta$ and $N$. When $k$ goes from $0$ to $N$, $\alpha$ will swing from $\theta$ to $\pi - \theta$ and vice versa. It is not hard to show that if $N (1+\frac{\zeta}{\pi}) \in \mathbb{Z} $, then different eigenvalues can occupy same points in each region (degeneracy happens) Fig. \ref{DegEig}. In fact the degeneracy condition is 
\begin{equation} \label{DegCond}
k + k ' =N (1+\frac{\zeta}{\pi})
\end{equation}

Using linearity of trace operator, one can rewrite Eq.\ref{LmtngDensityMat2} as
\begin{equation} \label{LmtngDensityMat3}
\tilde{\rho}  = \sum\limits_{k,k' = 0}^{N - 1} {\left| k \right\rangle \left\langle {k'} \right| \otimes } {{\Theta}\left( {k,k'} \right)}
\end{equation}
where
\begin{equation} \label{Theta}
{\Theta}\left( {k,k'} \right) = T{r_2}\left( {I \otimes \left| {{\psi _k}} \right\rangle \left\langle {{\psi _k{'}}} \right|M} \right)
\end{equation}
index 2 means that the trace takes over part 2 and
\begin{equation} \label{MMatrix}
M (k,k')= {\sideset {}{'}\sum\limits_{i = 0}^{1} {\left| {\lambda _k^{(i)}} \right\rangle \left\langle {\lambda _{k'}^{(i)}} \right| \otimes \left| {\lambda _{k'}^{(i)}} \right\rangle \left\langle {\lambda _k^{(i)}} \right|}}.
\end{equation}
It should be noticed that in $M$ the subscripts $k$ and $k'$ should satisfy the degeneracy condition ${\lambda _k^{(i)}} = {\lambda _{k'}^{(i)}}$.\\
$\tilde{\rho}$ in \eqref{LmtngDensityMat3} is general form of asymptotic average density matrix of QWC which can be used for calculating some important features such as asymptotic reduced density matrix $\tilde{\rho_{c}}$ or limiting distribution $\pi (v)$.\\
We have calculated a compact form of $\tilde{\rho_{c}}$ and $\pi(v)$ in the following corollaries:

\textbf{Corollary 1:} Suppose a QWC with $N$ nodes and a $U(2)$ coin operator
\begin{eqnarray} \label{GeneralCoin1}
\Gamma= e^{\frac{i \eta} {2}} \left[ {\begin{array}{*{20}{c}}
	{{e^{i\zeta }}\cos (\theta )}&{{e^{i\xi }}\sin (\theta )}\\
	{ - {e^{ - i\xi }}\sin (\theta )}&{{e^{ - i\zeta }}\cos (\theta )}
	\end{array}} \right]
\end{eqnarray}
The asymptotic reduced density matrix $\tilde{\rho_{c}}$ for initial state $\left| \chi_{0} \right\rangle$ is
\begin{equation}\label{rhoC}
\tilde{\rho_{c}}= \sum_{k=0}^{N-1} \Theta(k,k)
\end{equation}
where
\begin{equation} \label{Theta1}
{\Theta}\left( {k,k} \right) = T{r_2}\left( {I \otimes \left| {{\psi _k}} \right\rangle \left\langle {{\psi _{k}}} \right|M(k,k)} \right)
\end{equation}
with $\left| \psi_{k} \right\rangle = \left\langle k | \chi_{0} \right\rangle$ and the explicit form of M and the proof for \eqref{rhoC} are given in Appendix \ref{App1}.\\
$\tilde{\rho_{c}}$ can be used to calculate different parameters such as asymptotic entanglement temperature \cite{Diaz2016} and entanglement \cite{Carneiro2005}.

\textbf{Corollary 2:} Suppose a QWC with $N$ nodes and a $U(2)$ coin operator
\begin{eqnarray} \label{GeneralCoin2}
\Gamma= e^{\frac{i \eta} {2}} \left[ {\begin{array}{*{20}{c}}
	{{e^{i\zeta }}\cos (\theta )}&{{e^{i\xi }}\sin (\theta )}\\
	{ - {e^{ - i\xi }}\sin (\theta )}&{{e^{ - i\zeta }}\cos (\theta )}
	\end{array}} \right]
\end{eqnarray}
and $\left| \psi_{k} \right\rangle = \left\langle k | \chi_{0} \right\rangle$ is projection of initial state in $\left| k \right\rangle$ basis. The limiting distribution is given by
\begin{align}\label{pi(Xi)}
\pi \left( v \right) = \frac{1}{N} +\frac{1}{N} Re \left( \sum\limits_{\substack{	{k = 0}\\
 		{k \neq \frac{N \zeta}{2 \pi} } \\ {k \neq \frac{N \zeta}{2 \pi} + \frac{N}{2}}}}^{N - 1} {{e^{\frac{{2iv\left( {\frac{{2k}}{N} - \zeta } \right)}}{N}}}Tr\left( {{\Theta}\left( {k,\frac{{N\zeta }}{\pi } - k} \right)} \right)} \right)
\end{align}
where
\begin{equation} \label{Theta2}
{\Theta}\left( {k,k'} \right) = T{r_2}\left( {I \otimes \left| {{\psi _k}} \right\rangle \left\langle {{\psi _{k'}}} \right|M(k,k')} \right)
\end{equation}
and we exclude $k=\frac{N \zeta}{2 \pi} , \frac{N \zeta}{2 \pi} + \frac{N}{2}$ because they play no role in degeneracy of the system. The explicit form of $M(k,\frac{{N\zeta }}{\pi } - k)$ and the proof for \eqref{pi(Xi)} are given in Appendix \ref{App3}.

So the same as reduced density matrix, one just needs to know the constant matrix and subsequently the matrix ${{\Theta}\left( {k, {\frac{{N\zeta }}{\pi } - k}} \right)}$ to estimate limiting distribution for any coin and initial state, even for quantum walks with non-local initial states.

\section{\label{sec:level3}Examples}
In this section we are going to show that not only our formalism is more simple to work with but it is also more general and can be used to study more complicated cases.
\subsection{\label{sec:level4}Asymptotic Entanglement Temperature}
\begin{figure}
	\includegraphics[scale=.5]{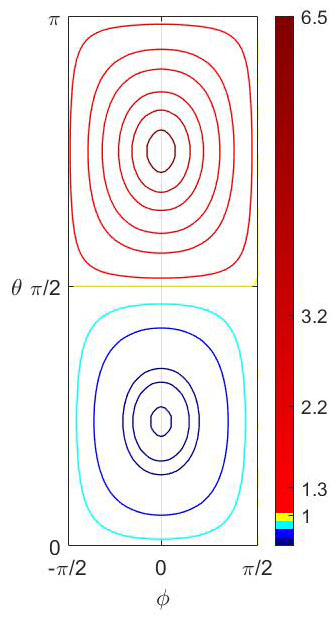}
	\caption{Isothermal curves for initial position \eqref{GenInitState} and $N=100$. Isotherms for hot zones are drawn as: $\frac{T}{T_{0}}=6.5, 3.2, 2.2, 1.6, 1.3, 1.06 $ and for cold zones, the isotherms are drawn as: $\frac{T}{T_{0}}=0.9, 0.8, 0.7, 0.68, 0.66$.}\label{TemperatureDiaz}
\end{figure}

By defining a thermodynamic equilibrium between position and chirality degrees of freedom, Romanelli \cite{Romanelli2016} defined a temperature concept. Diaz et.al. \cite{Diaz2016} analyzed asymptotic entanglement temperature for Hadamard QWCs with a general initial state.\\
Considering the general coin \eqref{GeneralCoin} and initial state as
\begin{equation}\label{GenInitState}
| \psi_{k} \rangle=\left[ \begin {array}{c} {\cos \left( \gamma/2 \right)}
\\ \noalign{\medskip}{{{\rm e}^{i\phi}}\sin \left( \gamma/2\right)}\end {array} \right]
\end{equation}
one can use $M(k,k)$ (App. \ref{App1}) with Hadamard coin ($\xi = \zeta = \frac{\pi}{2}$ and $\theta=\frac{\pi}{4}$)
\begin{equation}
\begin{split}
&M(k,k)= {\frac{1}{ 2 \left(\cos ^{2} \left( \omega \right)+1\right)}} \times \\
&{\left[ \begin {array}{cccc} 2\, \cos^{2} \left( \omega \right) +1&{{\rm e}^{-i\omega}}\cos \left( \omega \right) &{
	{\rm e}^{-i\omega}}\cos \left( \omega \right) &{{\rm e}^{-2\,i\omega}}
\\ \noalign{\medskip}{{\rm e}^{i\omega}}\cos \left( \omega \right) &1&
1&-{{\rm e}^{-i\omega}}\cos \left( \omega \right) 
\\ \noalign{\medskip}{{\rm e}^{i\omega}}\cos \left( \omega \right) &1&
1&-{{\rm e}^{-i\omega}}\cos \left( \omega \right) 
\\ \noalign{\medskip}{{\rm e}^{2\,i\omega}}&-{{\rm e}^{i\omega}}\cos
\left( \omega \right) &-{{\rm e}^{i\omega}}\cos \left( \omega
\right) &2\,  \cos^{2} \left( \omega \right) +1
\end {array} \right]}
\end{split}
\end{equation}
where $\omega = \frac{2 \pi k}{N}$. So by using \eqref{rhoC} and \eqref{Theta1} the explicit form of $\tilde{\rho}_{c}$ is easy to calculate. We have calculated eigenvalues of  $\tilde{\rho}_{C}$, i.e. $\lambda_{1}$ and $\lambda_{2}$ and using $T = \frac{{2{E_0}}}{{\ln \left( {\frac{{{\lambda _1}}}{{{\lambda _2}}}} \right)}}$ \cite{Diaz2016}, we plotted transient entanglement temperature for $N=100$ in \ref{TemperatureDiaz}. As you can see this is exactly the same as results provided by Diaz et.al. \cite{Diaz2016}.

We should note that the coin has been used in \cite{Diaz2016} is
\begin{equation}\label{DiazCoin}
C=\left[ {\begin{array}{*{20}{c}}
	{\cos \theta }&{\sin \theta }\\
	{\sin \theta }&{ - \cos \theta }
	\end{array}} \right]
\end{equation}
which is a special type of $U(2)$ coin with one parameter $\theta$, but in our formalism, we have the most general form of $U(2)$ with 3 parameters, which enables us to investigate some cases hard to study by the formalism of \cite{Diaz2016}.
For example, let's try to answer this question: \\

Is the hottest (coldest) point calculated for Hadamard coin in \cite{Diaz2016} the absolute maximum (minimum) in ET (entanglement temperature) or we can tune missing phase parameters ($\zeta$ and $\xi$) to reach warmer (colder) points?\\

\begin{figure}
	\includegraphics[scale=.55]{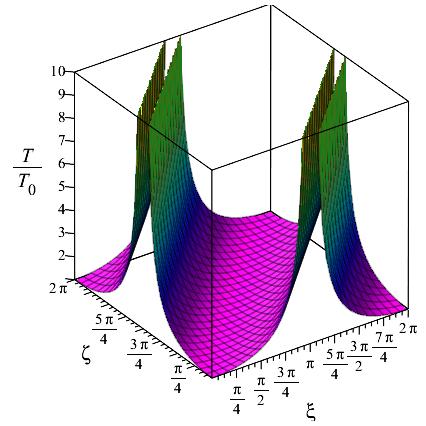}
	\caption{$\frac{T}{T_0}$ for hottest point $\left\langle {{\psi _k}} \right| = \frac{1}{{\sqrt N }}\left[ {\begin{array}{*{20}{c}}
			{\cos \frac{\pi }{8}}&{\sin \frac{\pi }{8}} \end{array}} \right]$ and $\theta = \frac{\pi}{4}$ versus $\zeta$ and $\xi$}\label{GenCoinHotPnt}
\end{figure}

\begin{figure}
	\includegraphics[scale=.4]{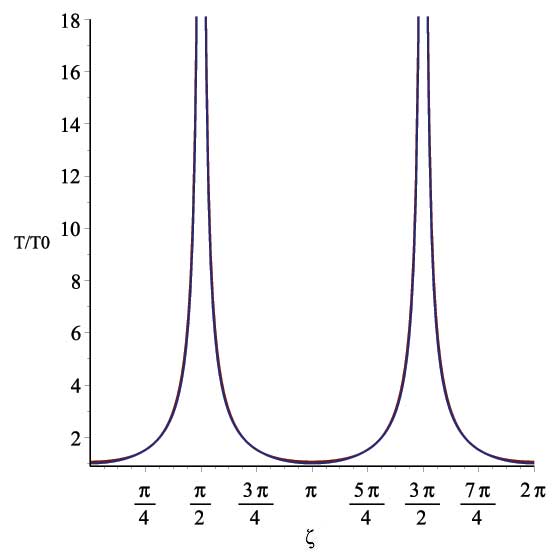}
	\caption{$\frac{T}{T_0}$ for hottest point $\left\langle {{\psi _k}} \right| = \frac{1}{{\sqrt N }}\left[ {\begin{array}{*{20}{c}}
			{\cos \frac{\pi }{8}}&{\sin \frac{\pi }{8}} \end{array}} \right]$ and $\theta = \frac{\pi}{4}$ versus $\zeta = \xi$}\label{GenCoinHotPntZ}
\end{figure}

The hottest point calculated in \cite{Diaz2016} for coin \eqref{DiazCoin} is $\left\langle {{\psi _k}} \right| = \left[ {\begin{array}{*{20}{c}}
	{\cos \frac{\pi }{8}}&{\sin \frac{\pi }{8}} \end{array}} \right]$. Using this initial state in \ref{Theta1} and putting $\theta = \frac{\pi}{4}$ in $M(k,k)$ in (App. \ref{App1}), the explicit form of $\tilde{\rho}_c$ is easy to calculate.
We have found eigenvalues of $\tilde{\rho}_c$ and plotted $\dfrac{T}{T_{0}}$ in Fig. \ref{GenCoinHotPnt} versus $(\zeta , \xi)$. Note that $T_0$ is the same as reference temperature used in \cite{Diaz2016}, i.e. ($\phi = 0$ and $\gamma = \pi$).
We found that for $\zeta = \xi$ there is no increase in temperature ($\dfrac{T}{T_0} = 1$), but in other cases we have a significant increasing. To see this dependency better, we have plotted the cross section view of Fig. \ref{GenCoinHotPnt} in ($\zeta = - \xi$) plane in Fig. \ref{GenCoinHotPntZ}. From Fig. \ref{GenCoinHotPntZ}, it is clear that, the minimum of $\frac{T}{T_{0}}$ is $1$, while the maximum can converge to infinity in certain values of $\zeta$ and $\xi$. So, tuning the parameters $\zeta$ and $\xi$ enables us to have more warmer points.

\begin{figure}
	\includegraphics[scale=.55]{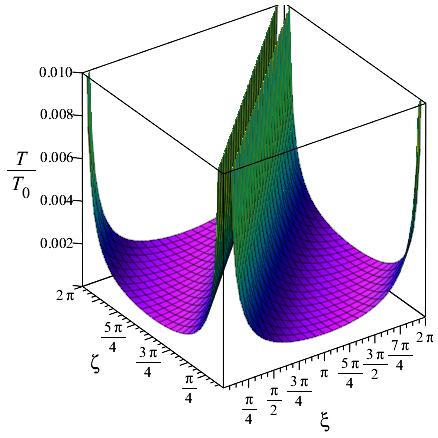}
	\caption{$\frac{T}{T_0}$ for coldest point $\left\langle {{\psi _k}} \right| = \frac{1}{{\sqrt N }}\left[ {\begin{array}{*{20}{c}}{\cos \frac{3\pi }{8}}&{\sin \frac{3\pi }{8}} \end{array}} \right]$ and $\theta = \frac{\pi}{4}$ versus $\zeta$ and $\xi$}\label{GenCoinCldPnt}
\end{figure}

Similar calculations show that by tuning $\zeta$ and $\xi$, colder points are accessible, see Fig. \ref{GenCoinCldPnt}.
The coldest point calculated in \cite{Diaz2016} for coin \eqref{DiazCoin} is $\left\langle {{\psi _k}} \right| = \left[ {\begin{array}{*{20}{c}}
	{\cos \frac{3 \pi }{8}}&{\sin \frac{3 \pi }{8}} \end{array}} \right]$. It is clear that for $\zeta = \xi$ there is no change in  $\frac{T}{T_0}$ in comparison to \cite{Diaz2016}, but in other cases there is a significant reduction in  $\frac{T}{T_0}$.

In summary, the parameters $\zeta$ and $\xi$ in \eqref{GeneralCoin} are important and should be taken into account, as $\zeta=\xi \pm \pi$ causes both hot points to get warmer and cold points to get colder.

\subsection{\label{sec:level5}Limiting Distribution with local initial state}

Limiting distribution (LD) for quantum walk on cycles has been studied widely and analytical solutions have been provided. Aharonov et.al. \cite{Aharonov2001} proved that the limiting distribution for quantum walk on cycles is uniform for odd number of nodes. For even number of nodes analytical solutions have been provided by Bednarska et.al. \cite{Bednarska2003} and Portugal \cite{Portugal2013}, however, the solutions are restricted to specific coins or initial states. But the \eqref{pi(Xi)} can be used to estimate limiting distribution for any coin and initial state.\\
For example lets find LD for Hadamard walk with initial state $\left| \psi_{0} \right \rangle= \left| 0 \right \rangle \otimes \left| 0 \right \rangle$, which is initial coin $\left| 0 \right \rangle$ localized at the origin ($x=0$). For using \eqref{pi(Xi)}, we need $\left| {{\psi _k}} \right\rangle \left\langle {{\psi _{k'}}} \right|$ and $M\left( {k,k'} \right)$ with $k' = \frac{{\zeta N}}{\pi } - k$ as
\begin{equation}
\left| {{\psi _k}} \right\rangle  = \left\langle k \right|\left. {{\psi _0}} \right\rangle = \frac{1}{{\sqrt N }}\left| 0 \right\rangle
\end{equation}
therefore
\begin{equation} \label{PsiPsi'}
\left| {{\psi _k}} \right\rangle \left\langle {{\psi _{k'}}} \right| = \frac{1}{N}\left| 0 \right\rangle \left\langle 0 \right|
\end{equation}
by putting \eqref{PsiPsi'} into \eqref{Theta2} and using \eqref{pi(Xi)}
\begin{equation}
\pi \left( v \right) = \frac{1}{N} + \frac{{{{\left( { - 1} \right)}^v}}}{{{N^2}}}\sum\limits_{{\substack{ k,k' = 0 \\ k \ne \frac{N}{4},\frac{3N}{4}}}}^{N - 1} {\frac{{\sin \left( {\frac{{2\pi k}}{N}} \right)\sin \left( {\frac{{2\pi k}}{N}\left( {2v - 1} \right)} \right)}}{{{{\cos }^2}\left( {\frac{{2\pi k}}{N}} \right) + 1}}}
\end{equation}
which is exactly the same as expressions derived by \cite{Bednarska2003} and \cite{Portugal2013}.
To see flexibility of our formalism lets find LD for initial coin $\left| 0 \right\rangle $ localized at $x=t$
\begin{equation}
\left| {{\psi _0}} \right\rangle  = \left| t \right\rangle  \otimes \left| 0 \right\rangle
\end{equation}
then
\begin{equation}
\left| {{\psi _k}} \right\rangle  = \left\langle k \right.\left| {{\psi _0}} \right\rangle  = \left\langle k \right.\left| t \right\rangle \left| 0 \right\rangle = \frac{1}{{\sqrt N }}{e^{\frac{{2i\pi kt}}{N}}}\left| 0 \right\rangle
\end{equation}
so
\begin{equation}
\left| {{\psi _k}} \right\rangle \left\langle {{\psi _{\frac{{\zeta N}}{\pi } - k}}} \right| = \frac{1}{N}{e^{ - 2i\left( {\frac{{2\pi k}}{N} - \zeta } \right)t}}\left| 0 \right\rangle \left\langle 0 \right|
\end{equation}
If we put it into \eqref{Theta2} and use \eqref{pi(Xi)}, we will have
\begin{equation}
\begin{array}{c}
\pi (v) = \frac{1}{N} + \frac{(-1)^{v-t}}{N^{2}} \sum\limits_{{\substack{ k,k' = 0 \\ k \ne \frac{N}{4},\frac{3N}{4}}}}^{N - 1} {\frac{\sin \left( 2\,{\frac {\pi \,k}{N}} \right) \sin \left({\frac{2 \pi \,k}{N} \left( 2 \left(v - t \right) -1 \right)} \right)}{ \cos^{2} \left({\frac {2 \pi \,k}{N}} \right) +1}}
\end{array}
\end{equation}

\subsection{\label{sec:level6}Limiting Distribution with non-local initial state}
\begin{figure}
	\includegraphics[scale=0.4]{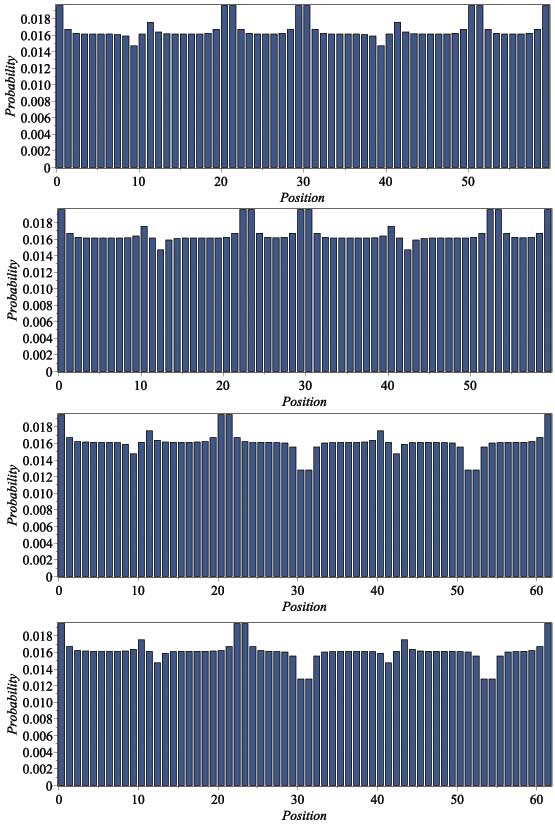}
	\caption{LD for entangled state, respectively from top to bottom: \\ 
			$\circ$ $N=60$ : $\left| {{\psi _0}} \right\rangle = \frac{1}{\sqrt{2}} (\left| {{0}} \right\rangle \left| {{0}} \right\rangle+ \left| {{20}} \right\rangle \left| {{1}} \right\rangle)$ \\ 
			$\circ$ $N=60$ : $\left| {{\psi _0}} \right\rangle = \frac{1}{\sqrt{2}} (\left| {{0}} \right\rangle \left| {{0}} \right\rangle+ \left| {{22}} \right\rangle \left| {{1}} \right\rangle)$ \\
			$\circ$ $N=62$ : $\left| {{\psi _0}} \right\rangle = \frac{1}{\sqrt{2}} (\left| {{0}} \right\rangle \left| {{0}} \right\rangle+ \left| {{20}} \right\rangle \left| {{1}} \right\rangle)$ \\
			$\circ$ $N=62$ : $\left| {{\psi _0}} \right\rangle = \frac{1}{\sqrt{2}} (\left| {{0}} \right\rangle \left| {{0}} \right\rangle+ \left| {{22}} \right\rangle \left| {{1}} \right\rangle)$ 
		}\label{EntFig}
\end{figure}

\begin{figure}
	\includegraphics[scale=0.4]{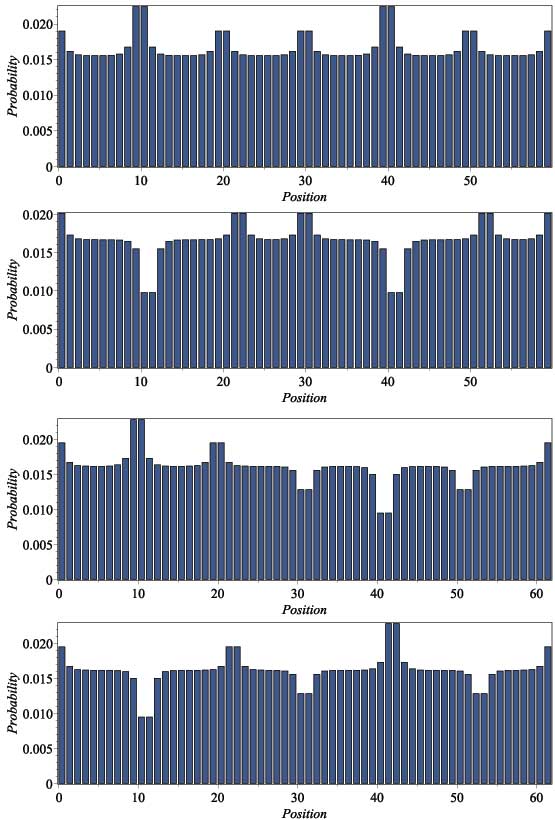}
	\caption{LD for separable state, respectively from top to bottom: \\ 
			$\circ$ $N=60$ : $\left| {{\psi _0}} \right\rangle = \frac{1}{\sqrt{2}} (\left| {{0}} \right\rangle \left| {{0}} \right\rangle+ \left| {{20}} \right\rangle \left| {{0}} \right\rangle)$ \\
			$\circ$ $N=60$ : $\left| {{\psi _0}} \right\rangle = \frac{1}{\sqrt{2}} (\left| {{0}} \right\rangle \left| {{0}} \right\rangle+ \left| {{22}} \right\rangle \left| {{0}} \right\rangle)$ \\
			$\circ$ $N=62$ : $\left| {{\psi _0}} \right\rangle = \frac{1}{\sqrt{2}} (\left| {{0}} \right\rangle \left| {{0}} \right\rangle+ \left| {{20}} \right\rangle \left| {{0}} \right\rangle)$ \\
			$\circ$ $N=62$ : $\left| {{\psi _0}} \right\rangle = \frac{1}{\sqrt{2}} (\left| {{0}} \right\rangle \left| {{0}} \right\rangle+ \left| {{22}} \right\rangle \left| {{0}} \right\rangle)$
		}\label{SepFig}
\end{figure}

Equation \eqref{pi(Xi)} is a general expression. So for non-local initial states, we just need to know $\left| {{\psi _k}} \right\rangle \left\langle {{{\psi}^{'} _k}} \right|$. For example, assume non-local entangled initial state as
\begin{equation}
\left| {{\psi _0}} \right\rangle = \frac{1}{\sqrt{2}} (\left| {{0}} \right\rangle \left| {{0}} \right\rangle+ \left| {{p}} \right\rangle \left| {{1}} \right\rangle)
\end{equation}
therefore

\begin{equation}
\begin{array}{l}
\left| {{\psi _k}} \right\rangle  = \left\langle {k}
\mathrel{\left | {\vphantom {k {{\psi _0}}}}
	\right. \kern-\nulldelimiterspace}
{{{\psi _0}}} \right\rangle  = \frac{1}{{\sqrt 2 }}\left( {\left\langle {k}
	\mathrel{\left | {\vphantom {k 0}}
		\right. \kern-\nulldelimiterspace}
	{0} \right\rangle \left| 0 \right\rangle  + \left\langle {k}
	\mathrel{\left | {\vphantom {k 1}}
		\right. \kern-\nulldelimiterspace}
	{1} \right\rangle \left| 1 \right\rangle } \right)\\
\,\,\,\,\,\,\,\,\,\,\,\,\,\, = \frac{1}{{\sqrt N }}\left( {\frac{{\left| 0 \right\rangle  + {e^{\frac{{2i\pi k}}{N}}}\left| 1 \right\rangle }}{{\sqrt 2 }}} \right)
\end{array}
\end{equation}
So
\begin{equation}
	\left| {{\psi _k}} \right\rangle \left\langle {{\psi _{\frac{{N\zeta }}{\pi } - k}}} \right| = \frac{1}{{2N}}\left( {\begin{array}{*{20}{c}}
		1&{{e^{2ip\left( {\frac{{\pi k}}{N} - \zeta } \right)}}}\\
		{{e^{2ip\left( {\frac{{\pi k}}{N}} \right)}}}&{{e^{2ip\left( {\frac{{2\pi k}}{N} - \zeta } \right)}}}
		\end{array}} \right)
\end{equation}
by plugging this into \eqref{pi(Xi)}, we will have limiting distribution (LD) for non-local entangled initial state. We have plotted LD for $p=20, 22$ in Fig. \ref{EntFig}. \\
In order to illustrate the role of entanglement in LD, we have plotted LD for separable initial state $\left| 0 \right\rangle \left| 0 \right\rangle  + \left| p \right\rangle \left| 0 \right\rangle$, i.e. coin $ \left| 0 \right\rangle $ is distributed in positions $0$ and $p$. Fig. \ref{SepFig} shows the LD for this separable initial state.

\section{\label{sec:level7}Conclusion}
We have used reduced density characteristic matrix (RDCM) approach introduced by \cite{Annabestani2019} to derive RDCM for QWC with general form of $U(2)$ coin operator (\textbf{Corollary 1}). We also showed that modified version of this approach can be used to derive an exact general expression for limiting distribution (\textbf{Corollary 2}) and it leads to same results provided in literatures (e.g. \cite{Bednarska2003}). Some features, such as entanglement temperature and limiting distribution have been plotted as examples of this approach, but investigating other coins and initial states will be simple.

\appendix

\section{Calculating $ \tilde{\rho_{c}} $} \label{App1}
The reduced density matrix $\tilde{\rho_{c}}$ is a result of tracing over the position subspace $x$. Since $\tilde{\rho_{c}}$ in \eqref{LmtngDensityMat3} is in k-space, we use completeness relation of $\sum_{x=1}^{N-1} \left| x \right\rangle \langle x|= I$ to change basis of $\tilde{\rho_{c}}$ from $\left| k \right\rangle$ to $\left| x \right\rangle$. So we can write
\begin{equation} \label{RedDensMatrixGenreal}
{{\tilde \rho }_c} = \sum\limits_{x = 1}^{N - 1} {T{r_p}\left( {\left( {\left| x \right\rangle \langle x| \otimes I} \right)\tilde{\rho} } \right)}
\end{equation}
By using \eqref{LmtngDensityMat3}
\begin{equation}
{{\tilde \rho }_c} = \sideset{}{'}\sum\limits_{k,k' = 0}^{\,\,\,\,\,\,\,\,\,N - 1\,\,'}  {{\delta _{k,k'}}{\Theta}\left( {k,k'} \right)} \\
\end{equation}
So,
\begin{equation}{{\tilde \rho }_c} = \sum\limits_{k = 0}^{N - 1} {{\Theta}\left( {k,k} \right)} 
\end{equation}
From \eqref{MMatrix}, one can see
\begin{equation} \label{Mkk}
M\left( {k,k} \right) = \frac{1}{a^2} \left[ \begin {array}{cccc} \frac{-b^2}{2}+{a}^{2}&-\overline{c}&-
\overline{c}&\frac{-b^2}{2}{{\rm e}^{-2\,i \left( \omega-\xi \right) 
	}}\\ \noalign{\medskip}-c&\frac{b^2}{2}&\frac{b^2}{2}&\overline{c}
	\\ \noalign{\medskip}-c&\frac{b^2}{2}&\frac{b^2}{2}&\overline{c}
	\\ \noalign{\medskip}\frac{-b^2}{2}{{\rm e}^{2\,i \left(\omega-\xi
			\right) }}&c&c&\frac{-b^2}{2}+{a}^{2}\end {array} \right]
\end{equation}
where
\begin{align}
\begin{split}
&a = \sin(\alpha)
\\
&b = \sin(\theta)
\\
&c = \frac{i}{2} b \sin \left( \omega - \zeta \right) \cos \left( \theta \right) {{\rm e}^{i \left( \omega-\xi \right) }}
\end{split}
\end{align}
with $\omega = \frac{2 \pi k}{N}$ and $ \cos(\alpha)= \cos(\theta) \cos(\omega - \zeta) $.

\section{Calculating general form of limiting distribution} \label{App3}

Using the probability distribution and eigenkets of the position space $| v \rangle $, one can estimate the probability of finding the particle in node $| v \rangle$. So using the limiting density matrix \eqref{LmtngDensityMat3}, the limiting probability at node $v$ can be given by
\begin{align} \label{LmtngDistRho}
\begin{split}
\pi \left( v \right) &= Tr\left( {\left( {\left| v \right\rangle \left\langle v \right|} \otimes I \right)\bar \rho } \right) \\
&= \sum\limits_{k,k' = 0}^{N - 1} {\left\langle {v}
	\mathrel{\left | {\vphantom {v k}}
		\right. \kern-\nulldelimiterspace}
	{k} \right\rangle \left\langle {{k'}}
	\mathrel{\left | {\vphantom {{k'} v}}
		\right. \kern-\nulldelimiterspace}
	{v} \right\rangle Tr\left( {{\Theta}\left( {k,k'} \right)} \right)}
\end{split}
\end{align}
which can be splitted into two summations as below
\begin{align} \label{PivSplit}
\pi \left( v \right) &= \sum\limits_{k = 0}^{N - 1} {\left\langle {v}
	\mathrel{\left | {\vphantom {v k}}
		\right. \kern-\nulldelimiterspace}
	{k} \right\rangle \left\langle {k}
	\mathrel{\left | {\vphantom {k v}}
		\right. \kern-\nulldelimiterspace}
	{v} \right\rangle Tr\left( {{\Theta}\left( {k,k} \right)} \right)} \nonumber \\
 &+ \sideset{}{'}\sum\limits_{\substack{ k,k' = 0  \\ k \ne k}}^{N - 1} {\left\langle {v}
	\mathrel{\left | {\vphantom {v k}}
		\right. \kern-\nulldelimiterspace}
	{k} \right\rangle \left\langle {{k'}}
	\mathrel{\left | {\vphantom {{k'} v}}
		\right. \kern-\nulldelimiterspace}
	{v} \right\rangle Tr\left( {{\Theta}\left( {k,k'} \right)} \right)} 
\end{align}
where the first summation is on $k=k'$ and the second summation just includes the terms resulted from degenerate cases \eqref{DegCond}, i.e. $k {'} = N (1 + \frac{\zeta}{\pi}) - k$. Using the Fourier transformation \eqref{Fourier} one can see in the first summation of \eqref{PivSplit} ${\left\langle {v} \mathrel{\left | {\vphantom {v k}} \right. \kern-\nulldelimiterspace} {k} \right\rangle \left\langle {k} \mathrel{\left | {\vphantom {k v}} \right. \kern-\nulldelimiterspace} {v} \right\rangle } = \frac{1}{N}$ and for the second summation ${\left\langle {v} \mathrel{\left | {\vphantom {v k}} \right. \kern-\nulldelimiterspace} {k} \right\rangle \left\langle {k'} \mathrel{\left | {\vphantom {k v}} \right. \kern-\nulldelimiterspace}{v} \right\rangle } = \frac{1}{N} {e^{\frac{{2i\pi v\left( {k - k'} \right)}}{N}}}$, so
\begin{align}\label{LmtngDist3}
\pi \left( v \right) &=\frac{1}{N} \sum\limits_{ k = 0}^{N - 1} Tr{{\Theta}\left( {k,k} \right)} \nonumber \\
&+\frac{1}{N} \sideset{}{'}\sum\limits_{\substack{ k,k' = 0 \\ k \ne k'}}^{N - 1}{{e^{\frac{{2i\pi v\left( {k-k'}\right)}}{N}}}Tr\left({{\Theta}\left( {k,k'} \right)} \right)} 
\end{align}
\\
To calculate the first term in \eqref{LmtngDist3} we substitute \eqref{MMatrix} into \eqref{Theta}, so we have
\begin{align}
\begin{split}
\begin{array}{l}
\sum\limits_{k = 0}^{N - 1} {Tr\left( {{\Theta}\left( {k,k} \right)} \right)}  = \sum\limits_{k = 0}^{N - 1} {Tr\left( {\sum\limits_{i = 0}^1 {\left| {\lambda _k^{(i)}} \right\rangle \langle \lambda _k^{(i)}|} } \right.} \\
\left. { \otimes \left\langle {{{\psi _k}}}
	\mathrel{\left | {\vphantom {{{\psi _k}} {\lambda _k^{(i)}}}}
		\right. \kern-\nulldelimiterspace}
	{{\lambda _k^{(i)}}} \right\rangle \left\langle {{\lambda _k^{(i)}}}
	\mathrel{\left | {\vphantom {{\lambda _k^{(i)}} {{\psi _k}}}}
		\right. \kern-\nulldelimiterspace}
	{{{\psi _k}}} \right\rangle } \right)
\end{array}
\end{split}
\end{align}
Now by applying the trace
\begin{align} \label{A2}
\begin{split}
&\sum\limits_{k = 0}^{N - 1} {Tr\left( {{\Theta}\left( {k,k'} \right)} \right)}  = \\ &\sum\limits_{k = 0}^{N - 1} {\sum\limits_{i = 0}^1 {\left\langle {{\lambda _k^{(i)}}}
		\mathrel{\left | {\vphantom {{\lambda _k^{(i)}} {\lambda _k^{(i)}}}}
			\right. \kern-\nulldelimiterspace}
		{{\lambda _k^{(i)}}} \right\rangle \left\langle {{{\psi _k}}}
		\mathrel{\left | {\vphantom {{{\psi _k}} {\lambda _k^{(i)}}}}
			\right. \kern-\nulldelimiterspace}
		{{\lambda _k^{(i)}}} \right\rangle \left\langle {{\lambda _k^{(i)}}}
		\mathrel{\left | {\vphantom {{\lambda _k^{(i)}} {{\psi _k}}}}
			\right. \kern-\nulldelimiterspace}
		{{{\psi _k}}} \right\rangle } } 
\end{split}
\end{align}
so,
\begin{equation}
\sum\limits_{k = 0}^{N - 1} {Tr\left( {{\Theta}\left( {k,k} \right)} \right)}  = \sum\limits_{k = 0}^{N - 1} {\left\langle {{\psi _k}} | {{\psi _k}} \right\rangle }
\end{equation}
where in this equation we use completeness relation $\sum\limits_{i = 0}^1 {\left| {\lambda _k^{(i)}} \right\rangle \left\langle {\lambda _k^{(i)}} \right|}  = I$ and ${\left\langle {{\lambda _k^{(i)}}}
	\mathrel{\left | {\vphantom {{\lambda _k^{(i)}} {\lambda _k^{(i)}}}}
		\right. \kern-\nulldelimiterspace}
	{{\lambda _k^{(i)}}} \right\rangle } =1$, so using \eqref{KdotPsi} 
\begin{align}
\sum\limits_{k = 0}^{N - 1} {Tr\left( {{\Theta}\left( {k,k} \right)} \right)}  = \sum\limits_{k = 0}^{N - 1} {\left\langle {{{\Psi _0}}}
	\mathrel{\left | {\vphantom {{{\Psi _0}} k}}
		\right. \kern-\nulldelimiterspace}
	{k} \right\rangle \left\langle {k}
	\mathrel{\left | {\vphantom {k {{\Psi _0}}}}
		\right. \kern-\nulldelimiterspace}
	{{{\Psi _0}}} \right\rangle } = \left\langle {{{\Psi _0}}}
\mathrel{\left | {\vphantom {{{\Psi _0}} {{\Psi _0}}}}
	\right. \kern-\nulldelimiterspace}
{{{\Psi _0}}} \right\rangle  = 1
\end{align}
For the second term in \eqref{LmtngDist3}, $\Sigma ^{'}$ means that the summation is just over the terms with ${\lambda _k^{(i)}} = {\lambda _{k'}^{(i)}}$ which are degenerate cases. But as we show in \eqref{DegCond}, $k' = N\left( {1 + \frac{\zeta }{\pi }} \right) - k$. By considering the fact that $k$ and $k'$ are labels which evolve in a cyclic manner, the degeneracy condition will be $k' = \left( {\frac{{N\zeta }}{\pi } - k} \right)\bmod N$. So, we just need to put $k' = \frac{{N\zeta }}{\pi } - k$ into \eqref{LmtngDist3}. Therefore
\begin{align}\label{pi(Xi)App}
\pi \left( v \right) &= \frac{1}{N} \\ &+\frac{1}{N} Re \left( \sum\limits_{\begin{array}{*{20}{c}}
	{k = 0}\\
	{ k \neq \frac{N \zeta}{2 \pi},\frac{N \zeta}{2 \pi} + \frac{N}{2}}
	\end{array}}^{N - 1} {{e^{\frac{{2iv\left( {\frac{{2k}}{N} - \zeta } \right)}}{N}}}Tr\left( {{\Theta}\left( {k,\frac{{N\zeta }}{\pi } - k} \right)} \right)} \right)
\end{align}
As we can see, we need $ {{\Theta}\left( {k,\frac{{N\zeta }}{\pi } - k} \right)}$. So, we need $M ( k, {\frac{{N\zeta }}{\pi } - k} )$ too. It is not hard to show that
\begin{equation} \label{Mkk'}
\begin{split}
&M ( k, {\frac{{N\zeta }}{\pi } - k} )= \\ &\frac{1}{a}\left[ {\begin{array}{*{20}{c}}
	{{{\left| b \right|}^2}}&{bc}&{b{c^*}}&{ - {b^2}}\\
	{b{c^*}}&{{{\left| b \right|}^2}{e^{ - 2i\beta }}}&{a - {{\left| b \right|}^2}}&{b\cos \theta }\\
	{{{\left( {bc} \right)}^*}}&{a - {{\left| b \right|}^2}}&{{{\left| b \right|}^2}\left( {2{e^{i\beta }}\cos \beta  - 1} \right)}&{ - b{c^*}}\\
	{{{\left( { - {b^2}} \right)}^*}}&{ - {b^*}c}&{ - bc}&{{{\left| b \right|}^2}}
	\end{array}} \right]
\end{split}
\end{equation}
where
\begin{align}
	\begin{split}
		a &= 2{\sin ^2}\alpha \\
		b &= \sin \theta {e^{ - i\lambda }}\\
		c &= - i{e^{ - i (\omega - \zeta) }}\cos \theta \sin (\omega - \zeta).
	\end{split}
\end{align}


\bibliography{references}

\providecommand{\noopsort}[1]{}\providecommand{\singleletter}[1]{#1}%
\begin{thebibliography}{24}%
\makeatletter
\providecommand \@ifxundefined [1]{%
 \@ifx{#1\undefined}
}%
\providecommand \@ifnum [1]{%
 \ifnum #1\expandafter \@firstoftwo
 \else \expandafter \@secondoftwo
 \fi
}%
\providecommand \@ifx [1]{%
 \ifx #1\expandafter \@firstoftwo
 \else \expandafter \@secondoftwo
 \fi
}%
\providecommand \natexlab [1]{#1}%
\providecommand \enquote  [1]{``#1''}%
\providecommand \bibnamefont  [1]{#1}%
\providecommand \bibfnamefont [1]{#1}%
\providecommand \citenamefont [1]{#1}%
\providecommand \href@noop [0]{\@secondoftwo}%
\providecommand \href [0]{\begingroup \@sanitize@url \@href}%
\providecommand \@href[1]{\@@startlink{#1}\@@href}%
\providecommand \@@href[1]{\endgroup#1\@@endlink}%
\providecommand \@sanitize@url [0]{\catcode `\\12\catcode `\$12\catcode
  `\&12\catcode `\#12\catcode `\^12\catcode `\_12\catcode `\%12\relax}%
\providecommand \@@startlink[1]{}%
\providecommand \@@endlink[0]{}%
\providecommand \url  [0]{\begingroup\@sanitize@url \@url }%
\providecommand \@url [1]{\endgroup\@href {#1}{\urlprefix }}%
\providecommand \urlprefix  [0]{URL }%
\providecommand \Eprint [0]{\href }%
\providecommand \doibase [0]{http://dx.doi.org/}%
\providecommand \selectlanguage [0]{\@gobble}%
\providecommand \bibinfo  [0]{\@secondoftwo}%
\providecommand \bibfield  [0]{\@secondoftwo}%
\providecommand \translation [1]{[#1]}%
\providecommand \BibitemOpen [0]{}%
\providecommand \bibitemStop [0]{}%
\providecommand \bibitemNoStop [0]{.\EOS\space}%
\providecommand \EOS [0]{\spacefactor3000\relax}%
\providecommand \BibitemShut  [1]{\csname bibitem#1\endcsname}%
\let\auto@bib@innerbib\@empty
\bibitem [{\citenamefont {Aharonov}\ \emph {et~al.}(1993)\citenamefont
  {Aharonov}, \citenamefont {Davidovich},\ and\ \citenamefont
  {Zagury}}]{YAharonov1993}%
  \BibitemOpen
  \bibfield  {author} {\bibinfo {author} {\bibfnamefont {Y.}~\bibnamefont
  {Aharonov}}, \bibinfo {author} {\bibfnamefont {L.}~\bibnamefont
  {Davidovich}}, \ and\ \bibinfo {author} {\bibfnamefont {N.}~\bibnamefont
  {Zagury}},\ }\href {\doibase 10.1103/PhysRevA.48.1687} {\bibfield  {journal}
  {\bibinfo  {journal} {Phys. Rev. A}\ }\textbf {\bibinfo {volume} {48}},\
  \bibinfo {pages} {1687} (\bibinfo {year} {1993})}\BibitemShut {NoStop}%
\bibitem [{\citenamefont {Ambainis}(2007)}]{Ambainis2007}%
  \BibitemOpen
  \bibfield  {author} {\bibinfo {author} {\bibfnamefont {A.}~\bibnamefont
  {Ambainis}},\ }\href {\doibase 10.1137/S0097539705447311} {\bibfield
  {journal} {\bibinfo  {journal} {SIAM J. Comput.}\ }\textbf {\bibinfo {volume}
  {37}},\ \bibinfo {pages} {210} (\bibinfo {year} {2007})}\BibitemShut
  {NoStop}%
\bibitem [{\citenamefont {Buhrman}\ and\ \citenamefont
  {\v{S}palek}(2006)}]{Buhrman2006}%
  \BibitemOpen
  \bibfield  {author} {\bibinfo {author} {\bibfnamefont {H.}~\bibnamefont
  {Buhrman}}\ and\ \bibinfo {author} {\bibfnamefont {R.}~\bibnamefont
  {\v{S}palek}},\ }in\ \href
  {http://dl.acm.org/citation.cfm?id=1109557.1109654} {\emph {\bibinfo
  {booktitle} {Proceedings of the Seventeenth Annual ACM-SIAM Symposium on
  Discrete Algorithm}}},\ \bibinfo {series and number} {SODA '06}\ (\bibinfo
  {publisher} {Society for Industrial and Applied Mathematics},\ \bibinfo
  {address} {Philadelphia, PA, USA},\ \bibinfo {year} {2006})\ pp.\ \bibinfo
  {pages} {880--889}\BibitemShut {NoStop}%
\bibitem [{\citenamefont {Magniez}\ and\ \citenamefont
  {Nayak}(2005)}]{Magniez2005}%
  \BibitemOpen
  \bibfield  {author} {\bibinfo {author} {\bibfnamefont {F.}~\bibnamefont
  {Magniez}}\ and\ \bibinfo {author} {\bibfnamefont {A.}~\bibnamefont
  {Nayak}},\ }\enquote {\bibinfo {title} {Quantum complexity of testing group
  commutativity},}\ in\ \href {\doibase 10.1007/11523468_106} {\emph {\bibinfo
  {booktitle} {Automata, Languages and Programming: 32nd International
  Colloquium, ICALP 2005, Lisbon, Portugal, July 11-15, 2005. Proceedings}}},\
  \bibinfo {editor} {edited by\ \bibinfo {editor} {\bibfnamefont
  {L.}~\bibnamefont {Caires}}, \bibinfo {editor} {\bibfnamefont {G.~F.}\
  \bibnamefont {Italiano}}, \bibinfo {editor} {\bibfnamefont {L.}~\bibnamefont
  {Monteiro}}, \bibinfo {editor} {\bibfnamefont {C.}~\bibnamefont
  {Palamidessi}}, \ and\ \bibinfo {editor} {\bibfnamefont {M.}~\bibnamefont
  {Yung}}}\ (\bibinfo  {publisher} {Springer Berlin Heidelberg},\ \bibinfo
  {address} {Berlin, Heidelberg},\ \bibinfo {year} {2005})\ pp.\ \bibinfo
  {pages} {1312--1324}\BibitemShut {NoStop}%
\bibitem [{\citenamefont {Farhi}\ \emph {et~al.}(2008)\citenamefont {Farhi},
  \citenamefont {Goldstone},\ and\ \citenamefont {Gutmann}}]{Farhi2007}%
  \BibitemOpen
  \bibfield  {author} {\bibinfo {author} {\bibfnamefont {E.}~\bibnamefont
  {Farhi}}, \bibinfo {author} {\bibfnamefont {J.}~\bibnamefont {Goldstone}}, \
  and\ \bibinfo {author} {\bibfnamefont {S.}~\bibnamefont {Gutmann}},\ }\href
  {\doibase 10.4086/toc.2008.v004a008} {\bibfield  {journal} {\bibinfo
  {journal} {Theory of Computing}\ }\textbf {\bibinfo {volume} {4}},\ \bibinfo
  {pages} {169} (\bibinfo {year} {2008})}\BibitemShut {NoStop}%
\bibitem [{\citenamefont {Reichardt}(2009)}]{Reichardt2009}%
  \BibitemOpen
  \bibfield  {author} {\bibinfo {author} {\bibfnamefont {B.~W.}\ \bibnamefont
  {Reichardt}},\ }in\ \href {\doibase 10.1109/FOCS.2009.55} {\emph {\bibinfo
  {booktitle} {Proceedings of the 2009 50th Annual IEEE Symposium on
  Foundations of Computer Science}}},\ \bibinfo {series and number} {FOCS '09}\
  (\bibinfo  {publisher} {IEEE Computer Society},\ \bibinfo {address}
  {Washington, DC, USA},\ \bibinfo {year} {2009})\ pp.\ \bibinfo {pages}
  {544--551}\BibitemShut {NoStop}%
\bibitem [{\citenamefont {Watrous}(2001)}]{Watrous2001}%
  \BibitemOpen
  \bibfield  {author} {\bibinfo {author} {\bibfnamefont {J.}~\bibnamefont
  {Watrous}},\ }\href@noop {} {\bibfield  {journal} {\bibinfo  {journal}
  {Journal of Computer and System Sciences}\ }\textbf {\bibinfo {volume}
  {62}},\ \bibinfo {pages} {376} (\bibinfo {year} {2001})}\BibitemShut
  {NoStop}%
\bibitem [{\citenamefont {Farhi}\ and\ \citenamefont
  {Gutmann}(1998)}]{Farhi1998}%
  \BibitemOpen
  \bibfield  {author} {\bibinfo {author} {\bibfnamefont {E.}~\bibnamefont
  {Farhi}}\ and\ \bibinfo {author} {\bibfnamefont {S.}~\bibnamefont
  {Gutmann}},\ }\href@noop {} {\bibfield  {journal} {\bibinfo  {journal}
  {Phys.\ Rev. A}\ }\textbf {\bibinfo {volume} {58}},\ \bibinfo {pages} {915}
  (\bibinfo {year} {1998})}\BibitemShut {NoStop}%
\bibitem [{\citenamefont {Ambainis}\ \emph {et~al.}(2005)\citenamefont
  {Ambainis}, \citenamefont {Kempe},\ and\ \citenamefont
  {Rivosh}}]{Ambainis2005}%
  \BibitemOpen
  \bibfield  {author} {\bibinfo {author} {\bibfnamefont {A.}~\bibnamefont
  {Ambainis}}, \bibinfo {author} {\bibfnamefont {J.}~\bibnamefont {Kempe}}, \
  and\ \bibinfo {author} {\bibfnamefont {A.}~\bibnamefont {Rivosh}},\ }in\
  \href {http://dl.acm.org/citation.cfm?id=1070432.1070590} {\emph {\bibinfo
  {booktitle} {Proceedings of the Sixteenth Annual ACM-SIAM Symposium on
  Discrete Algorithms}}},\ \bibinfo {series and number} {SODA '05}\ (\bibinfo
  {publisher} {Society for Industrial and Applied Mathematics},\ \bibinfo
  {address} {Philadelphia, PA, USA},\ \bibinfo {year} {2005})\ pp.\ \bibinfo
  {pages} {1099--1108}\BibitemShut {NoStop}%
\bibitem [{\citenamefont {Ambainis}\ \emph {et~al.}(2001)\citenamefont
  {Ambainis}, \citenamefont {Bach}, \citenamefont {Nayak}, \citenamefont
  {Vishwanath},\ and\ \citenamefont {Watrous}}]{Ambainis2001}%
  \BibitemOpen
  \bibfield  {author} {\bibinfo {author} {\bibfnamefont {A.}~\bibnamefont
  {Ambainis}}, \bibinfo {author} {\bibfnamefont {E.}~\bibnamefont {Bach}},
  \bibinfo {author} {\bibfnamefont {A.}~\bibnamefont {Nayak}}, \bibinfo
  {author} {\bibfnamefont {A.}~\bibnamefont {Vishwanath}}, \ and\ \bibinfo
  {author} {\bibfnamefont {J.}~\bibnamefont {Watrous}},\ }in\ \href {\doibase
  10.1145/380752.380757} {\emph {\bibinfo {booktitle} {Proceedings of the
  Thirty-third Annual ACM Symposium on Theory of Computing}}},\ \bibinfo
  {series and number} {STOC '01}\ (\bibinfo  {publisher} {ACM},\ \bibinfo
  {address} {New York, NY, USA},\ \bibinfo {year} {2001})\ pp.\ \bibinfo
  {pages} {37--49}\BibitemShut {NoStop}%
\bibitem [{\citenamefont {Mackay}\ \emph {et~al.}(2002)\citenamefont {Mackay},
  \citenamefont {Bartlett}, \citenamefont {Stephenson},\ and\ \citenamefont
  {Sanders}}]{Mackay2002}%
  \BibitemOpen
  \bibfield  {author} {\bibinfo {author} {\bibfnamefont {T.~D.}\ \bibnamefont
  {Mackay}}, \bibinfo {author} {\bibfnamefont {S.~D.}\ \bibnamefont
  {Bartlett}}, \bibinfo {author} {\bibfnamefont {L.~T.}\ \bibnamefont
  {Stephenson}}, \ and\ \bibinfo {author} {\bibfnamefont {B.~C.}\ \bibnamefont
  {Sanders}},\ }\href {http://stacks.iop.org/0305-4470/35/i=12/a=304}
  {\bibfield  {journal} {\bibinfo  {journal} {Journal of Physics A:
  Mathematical and General}\ }\textbf {\bibinfo {volume} {35}},\ \bibinfo
  {pages} {2745} (\bibinfo {year} {2002})}\BibitemShut {NoStop}%
\bibitem [{\citenamefont {Moradi}\ and\ \citenamefont
  {Annabestani}(2017)}]{Moradi2017}%
  \BibitemOpen
  \bibfield  {author} {\bibinfo {author} {\bibfnamefont {M.}~\bibnamefont
  {Moradi}}\ and\ \bibinfo {author} {\bibfnamefont {M.}~\bibnamefont
  {Annabestani}},\ }\href {\doibase 10.1088/1751-8121/aa9796} {\bibfield
  {journal} {\bibinfo  {journal} {Journal of Physics A: Mathematical and
  Theoretical}\ }\textbf {\bibinfo {volume} {50}},\ \bibinfo {pages} {505302}
  (\bibinfo {year} {2017})}\BibitemShut {NoStop}%
\bibitem [{\citenamefont {Moore}\ and\ \citenamefont
  {Russell}(2002)}]{Moore2002}%
  \BibitemOpen
  \bibfield  {author} {\bibinfo {author} {\bibfnamefont {C.}~\bibnamefont
  {Moore}}\ and\ \bibinfo {author} {\bibfnamefont {A.}~\bibnamefont
  {Russell}},\ }in\ \href {http://dl.acm.org/citation.cfm?id=646978.711707}
  {\emph {\bibinfo {booktitle} {Proceedings of the 6th International Workshop
  on Randomization and Approximation Techniques}}},\ \bibinfo {series and
  number} {RANDOM '02}\ (\bibinfo  {publisher} {Springer-Verlag},\ \bibinfo
  {address} {London, UK, UK},\ \bibinfo {year} {2002})\ pp.\ \bibinfo {pages}
  {164--178}\BibitemShut {NoStop}%
\bibitem [{\citenamefont {Aharonov}\ \emph {et~al.}(2001)\citenamefont
  {Aharonov}, \citenamefont {Ambainis}, \citenamefont {Kempe},\ and\
  \citenamefont {Vazirani}}]{Aharonov2001}%
  \BibitemOpen
  \bibfield  {author} {\bibinfo {author} {\bibfnamefont {D.}~\bibnamefont
  {Aharonov}}, \bibinfo {author} {\bibfnamefont {A.}~\bibnamefont {Ambainis}},
  \bibinfo {author} {\bibfnamefont {J.}~\bibnamefont {Kempe}}, \ and\ \bibinfo
  {author} {\bibfnamefont {U.}~\bibnamefont {Vazirani}},\ }in\ \href {\doibase
  10.1145/380752.380758} {\emph {\bibinfo {booktitle} {Proceedings of the
  Thirty-third Annual ACM Symposium on Theory of Computing}}},\ \bibinfo
  {series and number} {STOC '01}\ (\bibinfo  {publisher} {ACM},\ \bibinfo
  {address} {New York, NY, USA},\ \bibinfo {year} {2001})\ pp.\ \bibinfo
  {pages} {50--59}\BibitemShut {NoStop}%
\bibitem [{\citenamefont {Bednarska}\ \emph {et~al.}(2003)\citenamefont
  {Bednarska}, \citenamefont {Grudka}, \citenamefont {Kurzyński},
  \citenamefont {Łuczak},\ and\ \citenamefont {Wójcik}}]{Bednarska2003}%
  \BibitemOpen
  \bibfield  {author} {\bibinfo {author} {\bibfnamefont {M.}~\bibnamefont
  {Bednarska}}, \bibinfo {author} {\bibfnamefont {A.}~\bibnamefont {Grudka}},
  \bibinfo {author} {\bibfnamefont {P.}~\bibnamefont {Kurzyński}}, \bibinfo
  {author} {\bibfnamefont {T.}~\bibnamefont {Łuczak}}, \ and\ \bibinfo
  {author} {\bibfnamefont {A.}~\bibnamefont {Wójcik}},\ }\href {\doibase
  http://dx.doi.org/10.1016/j.physleta.2003.08.023} {\bibfield  {journal}
  {\bibinfo  {journal} {Physics Letters A}\ }\textbf {\bibinfo {volume}
  {317}},\ \bibinfo {pages} {21 } (\bibinfo {year} {2003})}\BibitemShut
  {NoStop}%
\bibitem [{\citenamefont {Kempe}(2005)}]{Kempe2005}%
  \BibitemOpen
  \bibfield  {author} {\bibinfo {author} {\bibfnamefont {J.}~\bibnamefont
  {Kempe}},\ }\href {\doibase 10.1007/s00440-004-0423-2} {\bibfield  {journal}
  {\bibinfo  {journal} {Probability Theory and Related Fields}\ }\textbf
  {\bibinfo {volume} {133}},\ \bibinfo {pages} {215} (\bibinfo {year}
  {2005})}\BibitemShut {NoStop}%
\bibitem [{\citenamefont {Abal}\ \emph {et~al.}(2006)\citenamefont {Abal},
  \citenamefont {Siri}, \citenamefont {Romanelli},\ and\ \citenamefont
  {Donangelo}}]{Abal2006}%
  \BibitemOpen
  \bibfield  {author} {\bibinfo {author} {\bibfnamefont {G.}~\bibnamefont
  {Abal}}, \bibinfo {author} {\bibfnamefont {R.}~\bibnamefont {Siri}}, \bibinfo
  {author} {\bibfnamefont {A.}~\bibnamefont {Romanelli}}, \ and\ \bibinfo
  {author} {\bibfnamefont {R.}~\bibnamefont {Donangelo}},\ }\href {\doibase
  10.1103/PhysRevA.73.042302} {\bibfield  {journal} {\bibinfo  {journal} {Phys.
  Rev. A}\ }\textbf {\bibinfo {volume} {73}},\ \bibinfo {pages} {042302}
  (\bibinfo {year} {2006})}\BibitemShut {NoStop}%
\bibitem [{\citenamefont {Annabestani}\ \emph {et~al.}(2010)\citenamefont
  {Annabestani}, \citenamefont {Akhtarshenas},\ and\ \citenamefont
  {Abolhassani}}]{Annabestani2010}%
  \BibitemOpen
  \bibfield  {author} {\bibinfo {author} {\bibfnamefont {M.}~\bibnamefont
  {Annabestani}}, \bibinfo {author} {\bibfnamefont {S.~J.}\ \bibnamefont
  {Akhtarshenas}}, \ and\ \bibinfo {author} {\bibfnamefont {M.~R.}\
  \bibnamefont {Abolhassani}},\ }\href {\doibase 10.1103/PhysRevA.81.032321}
  {\bibfield  {journal} {\bibinfo  {journal} {Phys. Rev. A}\ }\textbf {\bibinfo
  {volume} {81}},\ \bibinfo {pages} {032321} (\bibinfo {year}
  {2010})}\BibitemShut {NoStop}%
\bibitem [{\citenamefont {Romanelli}\ \emph {et~al.}(2014)\citenamefont
  {Romanelli}, \citenamefont {Donangelo}, \citenamefont {Portugal},\ and\
  \citenamefont {Marquezino}}]{Romanelli2014}%
  \BibitemOpen
  \bibfield  {author} {\bibinfo {author} {\bibfnamefont {A.}~\bibnamefont
  {Romanelli}}, \bibinfo {author} {\bibfnamefont {R.}~\bibnamefont
  {Donangelo}}, \bibinfo {author} {\bibfnamefont {R.}~\bibnamefont {Portugal}},
  \ and\ \bibinfo {author} {\bibfnamefont {F.~d.~L.}\ \bibnamefont
  {Marquezino}},\ }\href {\doibase 10.1103/PhysRevA.90.022329} {\bibfield
  {journal} {\bibinfo  {journal} {Phys. Rev. A}\ }\textbf {\bibinfo {volume}
  {90}},\ \bibinfo {pages} {022329} (\bibinfo {year} {2014})}\BibitemShut
  {NoStop}%
\bibitem [{\citenamefont {D\'{\i}az}\ \emph {et~al.}(2016)\citenamefont
  {D\'{\i}az}, \citenamefont {Donangelo}, \citenamefont {Portugal},\ and\
  \citenamefont {Romanelli}}]{Diaz2016}%
  \BibitemOpen
  \bibfield  {author} {\bibinfo {author} {\bibfnamefont {N.}~\bibnamefont
  {D\'{\i}az}}, \bibinfo {author} {\bibfnamefont {R.}~\bibnamefont
  {Donangelo}}, \bibinfo {author} {\bibfnamefont {R.}~\bibnamefont {Portugal}},
  \ and\ \bibinfo {author} {\bibfnamefont {A.}~\bibnamefont {Romanelli}},\
  }\href {\doibase 10.1103/PhysRevA.94.012305} {\bibfield  {journal} {\bibinfo
  {journal} {Phys. Rev. A}\ }\textbf {\bibinfo {volume} {94}},\ \bibinfo
  {pages} {012305} (\bibinfo {year} {2016})}\BibitemShut {NoStop}%
\bibitem [{\citenamefont {Annabestani}(2019)}]{Annabestani2019}%
  \BibitemOpen
  \bibfield  {author} {\bibinfo {author} {\bibfnamefont {M.}~\bibnamefont
  {Annabestani}},\ }\href {\doibase arXiv:1909.09214v1} {\  (\bibinfo {year}
  {2019}),\ arXiv:1909.09214v1}\BibitemShut {NoStop}%
\bibitem [{\citenamefont {Portugal}(2013)}]{Portugal2013}%
  \BibitemOpen
  \bibfield  {author} {\bibinfo {author} {\bibfnamefont {R.}~\bibnamefont
  {Portugal}},\ }\href@noop {} {\emph {\bibinfo {title} {Quantum Walks and
  Search Algorithms}}}\ (\bibinfo  {publisher} {Springer},\ \bibinfo {year}
  {2013})\BibitemShut {NoStop}%
\bibitem [{\citenamefont {Carneiro}\ \emph {et~al.}(2005)\citenamefont
  {Carneiro}, \citenamefont {Loo}, \citenamefont {Xu}, \citenamefont {Girerd},
  \citenamefont {Kendon},\ and\ \citenamefont {Knight}}]{Carneiro2005}%
  \BibitemOpen
  \bibfield  {author} {\bibinfo {author} {\bibfnamefont {I.}~\bibnamefont
  {Carneiro}}, \bibinfo {author} {\bibfnamefont {M.}~\bibnamefont {Loo}},
  \bibinfo {author} {\bibfnamefont {X.}~\bibnamefont {Xu}}, \bibinfo {author}
  {\bibfnamefont {M.}~\bibnamefont {Girerd}}, \bibinfo {author} {\bibfnamefont
  {V.}~\bibnamefont {Kendon}}, \ and\ \bibinfo {author} {\bibfnamefont {P.~L.}\
  \bibnamefont {Knight}},\ }\href {http://stacks.iop.org/1367-2630/7/i=1/a=156}
  {\bibfield  {journal} {\bibinfo  {journal} {New Journal of Physics}\ }\textbf
  {\bibinfo {volume} {7}},\ \bibinfo {pages} {156} (\bibinfo {year}
  {2005})}\BibitemShut {NoStop}%
\bibitem [{\citenamefont {Romanelli}(2012)}]{Romanelli2016}%
  \BibitemOpen
  \bibfield  {author} {\bibinfo {author} {\bibfnamefont {A.}~\bibnamefont
  {Romanelli}},\ }\href {\doibase 10.1103/PhysRevA.85.012319} {\bibfield
  {journal} {\bibinfo  {journal} {Phys. Rev. A}\ }\textbf {\bibinfo {volume}
  {85}},\ \bibinfo {pages} {012319} (\bibinfo {year} {2012})}\BibitemShut
  {NoStop}%
\end{thebibliography}%

\end{document}